\documentclass[12pt]{article} \usepackage{amsmath}
\usepackage{amssymb}
\usepackage{euscript}

\usepackage{epsfig}

\usepackage{cite}

\usepackage{alltt}

\usepackage{epic}
\begin{document}   \allowdisplaybreaks

%%%%%%%%%%%%%%%%%%%%%%%%%%%%%%%%%%%%%%%%%%%%%%%%%%%%%%%%%%%%%%%%%%%%%%%%%%%%%%%%%
%%%%%%%%%%%%%%%%%%%%%%%%%%%%%%%%%%%%%%%%%%%%%%%%%%%%%%%%%%%%%%%%%%%%%%%%%%%%%%%%%
\begin{titlepage}

\begin{flushright}
{\bf CERN-TH/2002-061} \\
{\bf  March 2002 }
\end{flushright}

\vspace{2mm}
\begin{center}{\bf\Large Spin polarization   }\end{center}
\begin{center}{\bf\Large and the Einstein--Podolsky--Rosen paradox  }\end{center}
\begin{center}{\bf\Large in the Monte Carlo event records $^{*}$}\end{center}
\vspace{1 cm}
\begin{center}
  {\large\bf   Z. W\c{a}s$^{a,b}$} 
\vspace{0.3 cm}
\\
{\em $^a$Institute of Nuclear Physics\\
         Kawiory 26a, 30-055 Cracow, Poland.}\\
{\em $^b$CERN, Theory Division, CH-1211 Geneva 23, Switzerland.}\\
\vspace{0.3 cm}
e-mail: {\tt Zbigniew.Was@cern.ch}\\
\vspace{1mm}

\end{center}

\vspace{1mm}
\begin{abstract}

In the future high energy physics experiments, the question of properly matching the 
phenomenological programs that describe different parts of the physics
processes (such as hard scattering, hadronization, decay of resonances, 
detector response, etc.) is very important. In  the past,
FORTAN common blocks filled with lists of objects (particles, strings, clusters, etc.)
of defined properties,
origins and descendants were in use. Similar structures are now envisaged, for future 
programs, to be written in languages such as C++ or Java.
From the physics point of view such an approach is not correct, since this kind of  
data structures impose certain approximations on the physics content.
In the present paper, we will explore their limits, using examples from the physics
of $W$'s, $\tau$'s and the Higgs boson, still to be discovered. 

\end{abstract}
\begin{center}
{\it Presented at CPP2001 on Automatic Calculation for Future Colliders \\
 Tokyo Metropolitan Univ., 28--30
November  2001.}
\end{center}

\vspace{3mm}
\begin{flushleft}
{\bf 
{\bf CERN-TH/2002-061} \\
  March  2002}
\end{flushleft}

\vspace{1mm}
\footnoterule
\noindent
{\footnotesize
\begin{itemize}
\item[${*}$]
This work is partly supported by
the Polish State Committee for Scientific Research 
(KBN) grant No. 
%5P03B12420, %%%<-- Z Pana ostatniej pracy
5P03B09320, %%%<-- Marek ... Z Pana ostatniej pracy
%5P03B10121, %%%<-- Gosia
and the European Commission 5th framework contract 
HPRN-CT-2000-00149.%%%<-- Z Pana ostatniej pracy
%2P03B11819, %%%<-- Ela
%KBN 2P03B08414, %%%<-- Marek
%the US DoE contracts DE-FG05-91ER40627 and DE-AC03-76SF00515,
%the Maria Sk\l{}odowska-Curie Joint Fund II PAA/DOE-97-316,
%and the Polish--French Collaboration within IN2P3 through LAPP Annecy.
\item[${\;}$] Home page: {\tt http://wasm.home.cern.ch/wasm/}
\end{itemize}
}

\end{titlepage}
%%%%%%%%%%%%%%%%%%%%%%%%%%%%%%%%%%%%%%%%%%%%%%%%%%%%%%%%%%%%%%%%%%%%%%%%%%%%%%%%%%%%%%%%%%%%%
%%%%%%%%%%%%%%%%%%%%%%%%%%%%%%%%%%%%%%%%%%%%%%%%%%%%%%%%%%%%%%%%%%%%%%%%%%%%%%%%%%%%%%%%%%%%%
%\section{Introduction}
%%%%%%%%%%%%%%%%%%%%%%%%%%%%%%%%%%%%%%%%%%%%%%%%%%%%%%%%%%%%%%%%%%%%%%%%%%%%%%%%%%%%%%%%%%%%%

At present, intensive studies are being performed to design future software architectures for
experiments on
proton proton colliders, such as the Tevatron \cite{Tevatron}  or the LHC \cite{LHC1,LHC2}
and high energy $e^+e^-$ linear colliders such as   JLC \cite{Abe:2001gc},
NLC \cite{:2001ve} or TESLA \cite{TESLA}. 

One of the important ingredients in such designs is the data structure for storing the
Monte Carlo events. It is generally accepted that the data structures based on objects
such as particles, clusters, strings, etc.
with properties such as tracks, momenta, colour, spin, mass, etc. and on the relations
explaining the origins and descendants of the objects is the most convenient one.
This is the case at present \cite{PDG:1998}, and it is also envisaged for the future, see 
\cite{Boos:2001cv}. At the same time such a picture is in conflict with the basic 
principles of quantum mechanics. Einstein--Rosen--Podolsky paradox is an example of such
phenomena. A general problem is that the quantum state of a multiparticle
system cannot (at least in principle) be represented as a statistical combination of the states
defined by the products of the pure quantum states of the individual particles.
It is thus of the utmost importance to examine whether
the approximation enforced by  the data structure is purely
academic, or if it rather represents a real difficulty, which may affect the interpretation
of the  future data.

It would not be a serious problem if the predictions of the Standard Model used in the 
interpretation
of the future data could be provided  by a single program, black box, without any need 
of analysing its parts. Then anything that would be measured beyond the prediction of such a
hypothetical Monte Carlo program%
\footnote{ As experimental data are always obtained with imperfect detectors,
cuts, inefficiencies, etc. theoretical predictions
must be convoluted with the experimental effects. 
Monte Carlo 
simulation techniques are the only tools, for the time being, able to complete the task.
 } 
would be interpreted as  ``new physics''. Agreement, on the other hand, would constitute
confirmation of the Standard Model, as it is understood at present (and proper functioning 
of the detector as well).

Because of the complexity of the problem, Monte Carlo predictions need to be 
dealt with by programs describing: the action of the detector and of the analysis, 
on the experimental side, and various effects, such as those from hard processes, 
hadronization, decay of resonances, etc.,  on the theoretical side. 
Every part is inevitably calculated 
with some approximation and, as a consequence, some systematic errors affect these
predictions.

In the following, we will omit   these complex issues from the discussion. We will limit ourselves 
to the question of spin effects, more precisely to the consequences of approximations
used in combining production and decays of the intermediate states. As examples
we will use effects in the production of pairs of $\tau$-leptons and $W$-bosons\footnote{%
For the sake of convenience, I will use as examples theoretical calculations
 I was involved in myself.}.

I advocate here that spin effects are non-treatable
in the scheme where properties are attributed to individual particles only,
in spite of the fact that 
it is  the very method we used in {\tt KORALZ} \cite{koralz4:1994} --
the program widely used 
at LEP for the simulation  of $\tau$-lepton pair production 
and decay, including spin and QED bremssstrahlung effects. 
As described in ref. \cite{koralz4:1994} the algorithm 
of spin generation for  any individual event
 was consisting of the following steps:

\begin{enumerate}
\item
 An event consisting of  a pair of $\tau$ leptons, bremsstrahlung photons,
etc., was generated. 
\item
 Helicity states for both $\tau^+$ and $\tau^-$ were generated.
At this point, an   approximation with respect to quantum mechanisc was introduced.
\item
 Information on these helicty states, including the definition of quantization frames,
i.e. the relation between $\tau$'s rest frame and laboratory frame, was then  transmitted 
to {\tt TAUOLA} \cite{Jadach:1993hs,Jezabek:1991qp,Jadach:1990mz}, 
the package for the generation of $\tau$-lepton decays. 
\item
Finally  {\tt TAUOLA} performed decays of 100\% polarized $\tau$'s, and the event 
in the {\tt HEPEVT} common block was completed. 
It was not considered necessary to store the 
information about spin degrees of freedom;
however,   it proved convenient, for applications
that  rely on the approximate spin picture. 
\end{enumerate}

 At LEP, in (nearly) all cases, such an approach was sufficient. Thanks to the 
ultrarelativistic nature of $\tau$-leptons $( {m_\tau \over m_Z})^2
\ll 1$, missing effects were in most cases negligible. Let us note, however, that it 
was {\em not always} the case. Thanks to the excellent performance of the LEP detectors
it was necessary to revisit the complete spin effects \cite{Sanchez:1995gj}
and indeed the effects turned out to be measurable  \cite{Abreu:1997vp,Barate:1997mz}.  
Even more important
was the case of complete spin effects in the measurement of the $\tau$-lepton
lifetime, using the method of impact parameter sum (see e.g. \cite{Ferrante:1995da}).
In that case, terms missing in {\tt KORALZ} were not at all suppressed by the
mass factors. Fortunately we could recall the  solution of {\tt KORALB }
\cite{koralb:1985,jadach-was:1984}, which was always serving as a {\it backup solution} for 
the spin treatment in {\tt KORALZ}. {\tt KORALB} relies on the full spin  
density matrix, but includes first order bremsstrahlung 
corrections only; it also misses electroweak corrections,  necessary 
in high precision studies of LEP. 

We conclude that the solution for the spin treatment of $\tau$ leptons at LEP  
was optimal. On one side, a convenient picture
of particles with properties, origins and descendants could be used 
and, on the other, a complete  full spin solution was available, if necessary.

% convenient for most of the applications

Let us now turn to another example of the spin effect, this time in  the process, which
 can  be a source of a background.  
Let us consider a semi-realistic observable of invariant mass distribution 
for  pairs of $s$-flavoured jets, and its background from the four-jet process in
 the $e^+e^-$ annihilation into  four quarks
($ c\bar c s \bar s$ jets) at 
the 350 GeV centre-of-mass energy, with the  veto cut on $c$-jets forcing them
into directions close to the beam pipe. We will not discuss details of the study,
 which is presented
in ref. \cite{Ishikawa:1998ma}.
We will simply recall some numerical results from that paper.
In fig.~1,
%\ref{figj} , a 
thin line represents the result of the
complete {\tt CC-43}%
\footnote{
At the Born level 
there are in total {\bf 43} diagrams for the {\bf c}harge  {\bf c}urrent mediated
$e^+e^- \to c\bar s \bar c s$ process. This number is reduced to 
{\bf 3} if only the doubly $W$-resonant diagrams
are taken.
}
 matrix element. Not only the expected peak of $Z$-boson is
clearly visible, but there is another one, at high energies as well.
If we reduce the matrix element to the simple case of  {\tt CC-03}, where
the double resonant $WW$-pair production  is kept only  (thick line), 
the $Z$ resonance disappears, but the second peak remains visible. In fig.  \ref{fig2} we 
investigate the origin of the second peak even further.
We compare the {\tt CC-03} (distribution identical  to that in the previous figure) with the
case when transverse spin correlations are switched off. The difference is enormous.
Not only is the peak at high energies reduced by a factor of 4, but also
a huge shoulder of the distribution forms at lower invariant masses.
There is no question, if an observable like ours was used in a search 
for  new particles, that the lack of  proper spin effects in the code simulating the background 
could lead to difficulties in data analysis and even to the temporary ``discovery'' 
of non--existing resonance
for small  experimental samples.
%%%%%%%%%%%%%%%%%%%%%%%%%%%%%
\begin{figure}[!ht]
%\centering
\setlength{\unitlength}{1mm}
\begin{picture}(100,90)
\put( -10,-90){\makebox(0,0)[lb]{\epsfig{file=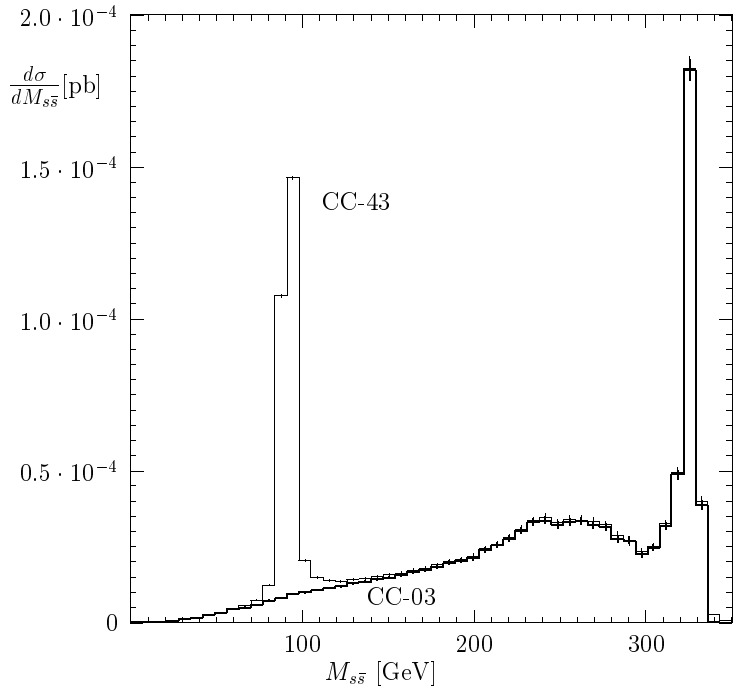,width=140mm,height=200mm}}}
\end{picture}
\begin{center}
\caption{
The ${d\sigma \over d M_{s\bar s}}$ differential
  distribution of the invariant mass of the 
``visible'' $s \bar s$ jet pair. Veto cut on  $c \bar c$ jets is applied. 
The centre-of-mass energy is 
350 GeV. Matrix element of type CC-03 (thick line)
 and type  CC-43 (thin line). 
For details, see the text and ref.~\cite{Ishikawa:1998ma}.
}
\end{center}
\label{figj}
\end{figure}
%%%%%%%%%%%%%%%%%%%%%%%%%%%%%
%%%%%%%%%%%%%%%%%%%%%%%%%%%%%
\begin{figure}[!ht]
%\centering
\begin{center}
\setlength{\unitlength}{1mm}
\begin{picture}(100,90)
\put( -50,-90){\makebox(0,0)[lb]{\epsfig{file=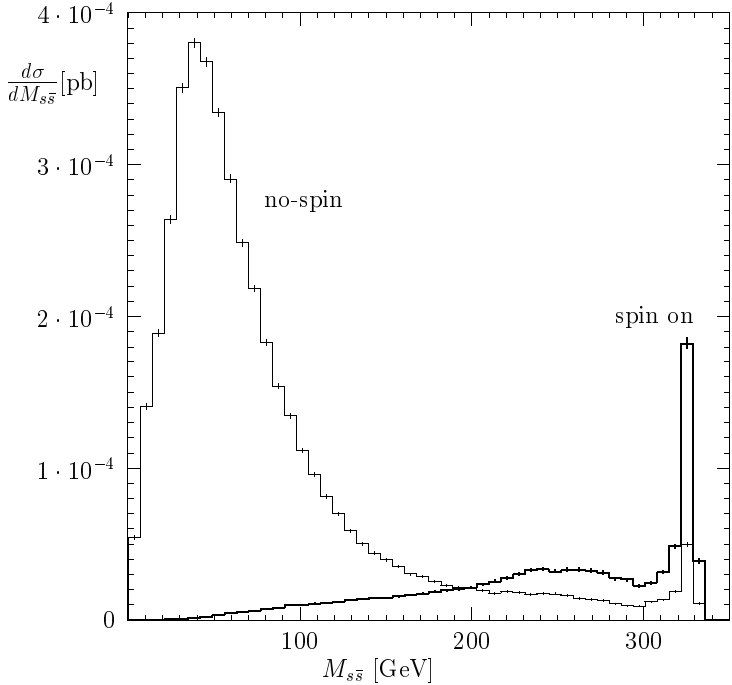,width=140mm,height=200mm}}}
\end{picture}
\end{center}
\caption{
The ${d\sigma \over d M_{s\bar s}}$ differential
  distribution of the invariant mass of the 
``visible'' $s \bar s$ jet pair. Veto cut on  $c \bar c$ jets is applied. 
The centre-of-mass energy is 
350 GeV. Matrix element of type CC-03: no spin correlation (thin line),
of  type  CC-03: with spin correlations switched on (thick line). 
For details, see the text and ref.~\cite{Ishikawa:1998ma}.
}
\label{fig2}
\end{figure}
%%%%%%%%%%%%%%%%%%%%%%%%%%%%%

Is this example really worrisome? Probably not. At the  
future high energy experiments, data will be collected in sufficient quantities, 
all  Monte Carlo 
programs will use  the matrix elements for the combined production
{\it and} decay of the $W$-pairs with  no approximations. 
Already now the physics of the $W$-pair
production is well established and the separation between production and decay 
is neither necessary (all its decay channels can be described by the same matrix element) 
nor convenient; effects due to the structure of the $W$ propagator make
the separation into production and decay  complicated.

Let us now turn to another example of the spin implementation algorithm.
It is taken from  ref.\cite{Pierzchala:2001gc}. The 
algorithm, essentially that of {\tt KORALZ}, was adopted to work with any Monte Carlo 
program providing the production of $\tau$-leptons. If the generated events are 
stored  in the format of a {\tt HEPEVT} common block, then  the  algorithm 
consisting  of the following basic steps can be used:
\begin{enumerate}
\item
Search for   $\tau$-leptons  in  a {\tt HEPEVT } common block 
(filled by any MC program).
\item
Check what the  origin of $\tau$--lepton is: 
$Z,\gamma,W,h,H^{\pm}$ or eventually, $2\to 2$--body process such as: 
  $e^+e^-,\; (u \bar u),\;  (d \bar d) \to \tau^+ \tau^-$.
\item
For the $2\to 2$--body process of $\tau$-pair production, it is sometimes
possible to calculate the $\tau$ polarization as a function of the invariant mass 
of the $\tau$--lepton pair and angle between the directions of  $\tau$--leptons and
incoming effective beams (in the rest frame of $\tau$-pair).
\item
If in addition to  the $\tau$-leptons,  photons
or partons (gluons, quarks, etc.) are stored in {\tt HEPEVT} common block,
one needs to define the ``effective incoming beams''.
\item
From such an information one can generate $\tau$ helicity states and define the
relation between the $\tau$ rest frame and the laboratory frame.
\item
The $\tau$ decay is generated with the help of a program such as {\tt TAUOLA}. 
\item
Finally the entire event  stored in a {\tt HEPEVT } common block is appended 
with the  $\tau$'s  decay products.
\item
Optionally the final-state bremsstrahlung (emission from $\tau$-leptons)  
can be generated using  {\tt PHOTOS} \cite{Barberio:1990ms,Barberio:1994qi}.
\end{enumerate}

\begin{figure}[!ht]
\begin{center}
\setlength{\unitlength}{1mm}
\begin{picture}(100,90)
\put(-50,-30){\makebox(0,0)[lb]{\epsfig{file=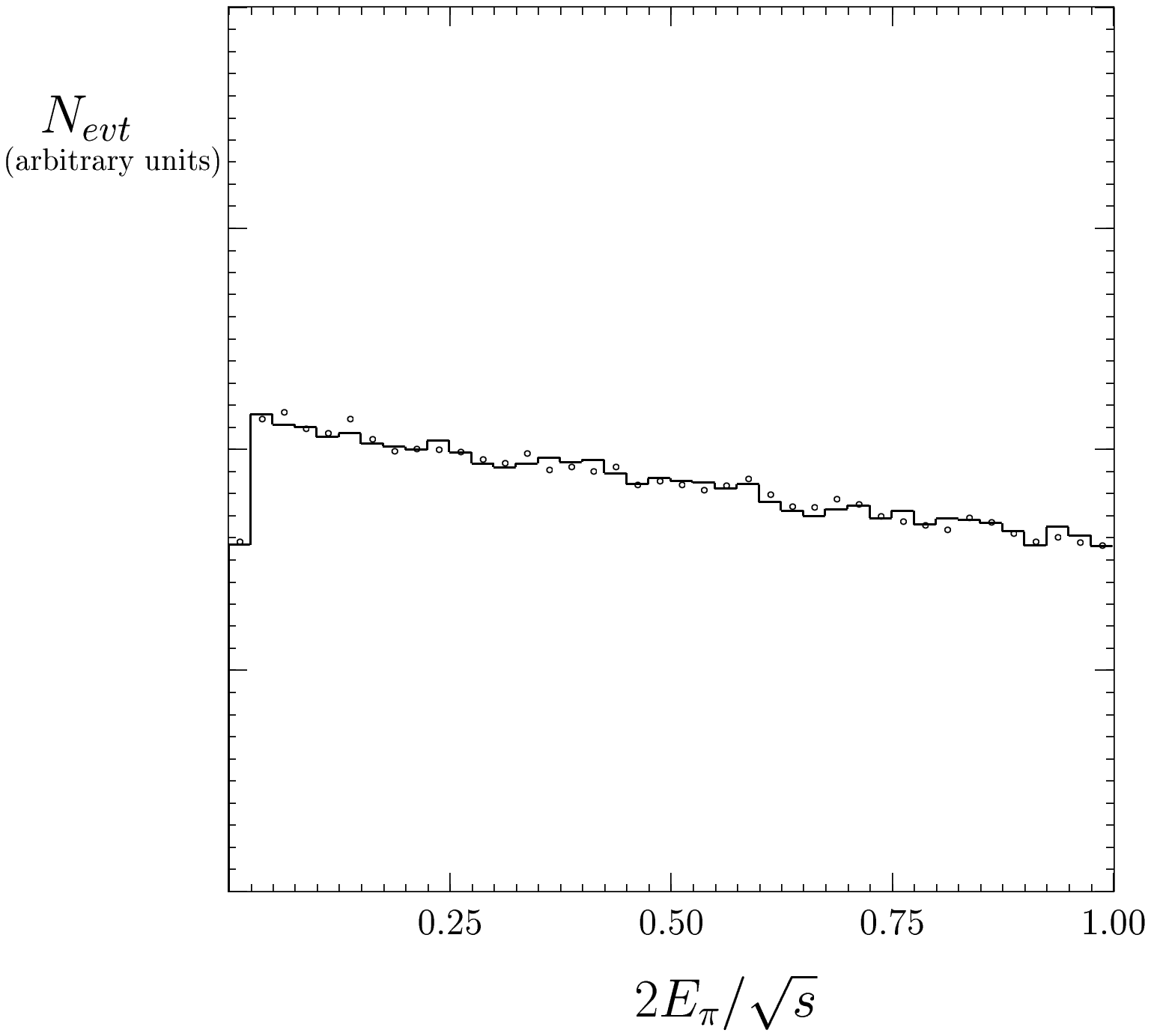,width=110mm,height=130mm}}}
\put( 30,-30){\makebox(0,0)[lb]{\epsfig{file=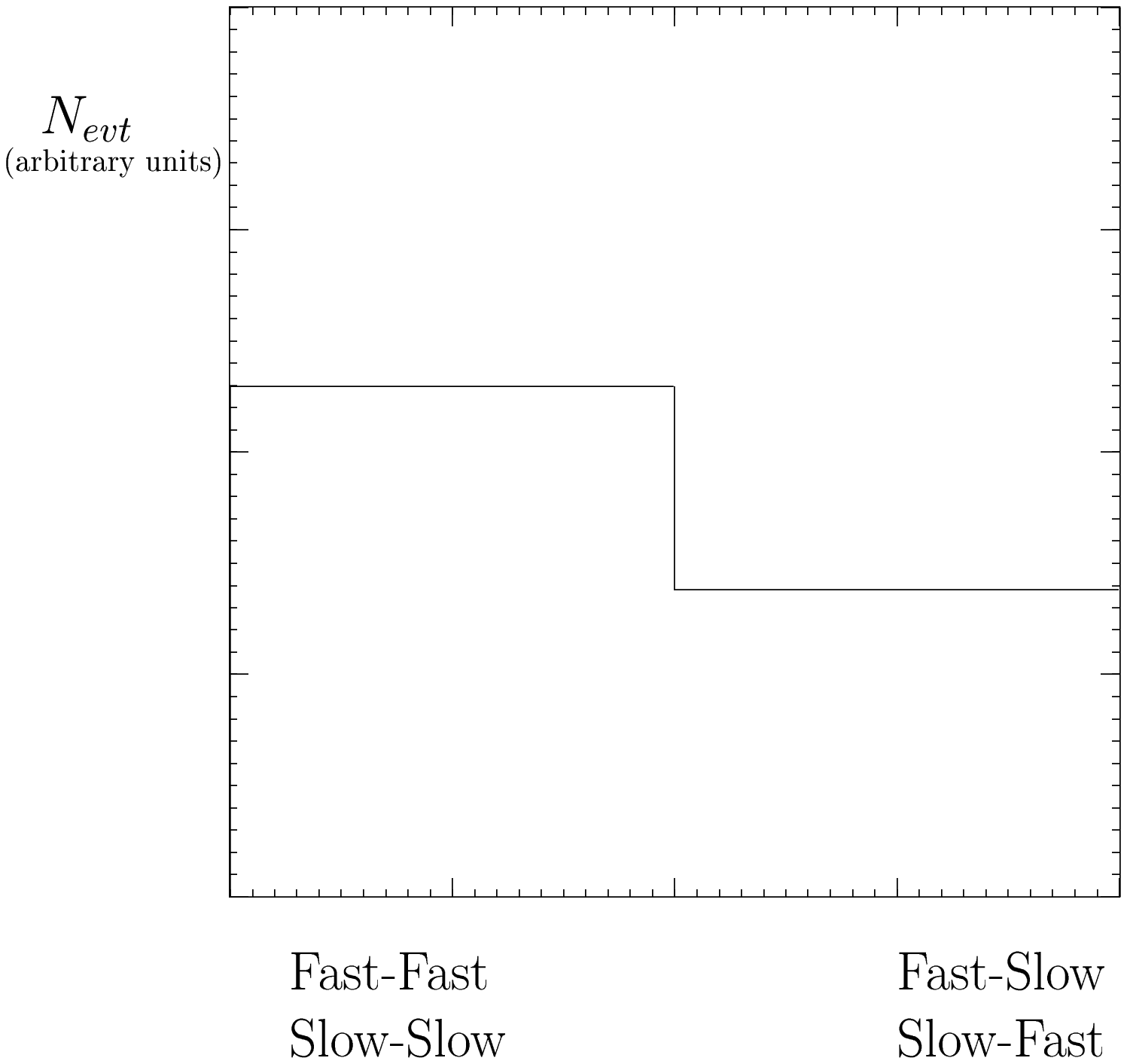,width=110mm,height=130mm}}}
\end{picture}
\end{center}
\caption{Basic properties of the spin effects in the case of $\tau$-leptons produced from 
$Z/\gamma$ intermediate state. The effects of  $\tau$ polarization are in the 
left-hand side plot.
We can see the slope of the $\pi^\pm$ energy slopes calculated in the $Z$ boson rest frame.
The effects of the spin correlations are in the right-hand side plot.
See ref.\cite{Pierzchala:2001gc} for details. }
\label{fig3}
\end{figure}

\begin{figure}[!ht]
\begin{center}
\setlength{\unitlength}{1mm}
\begin{picture}(100,90)
\put(-60,-30){\makebox(0,0)[lb]{\epsfig{file=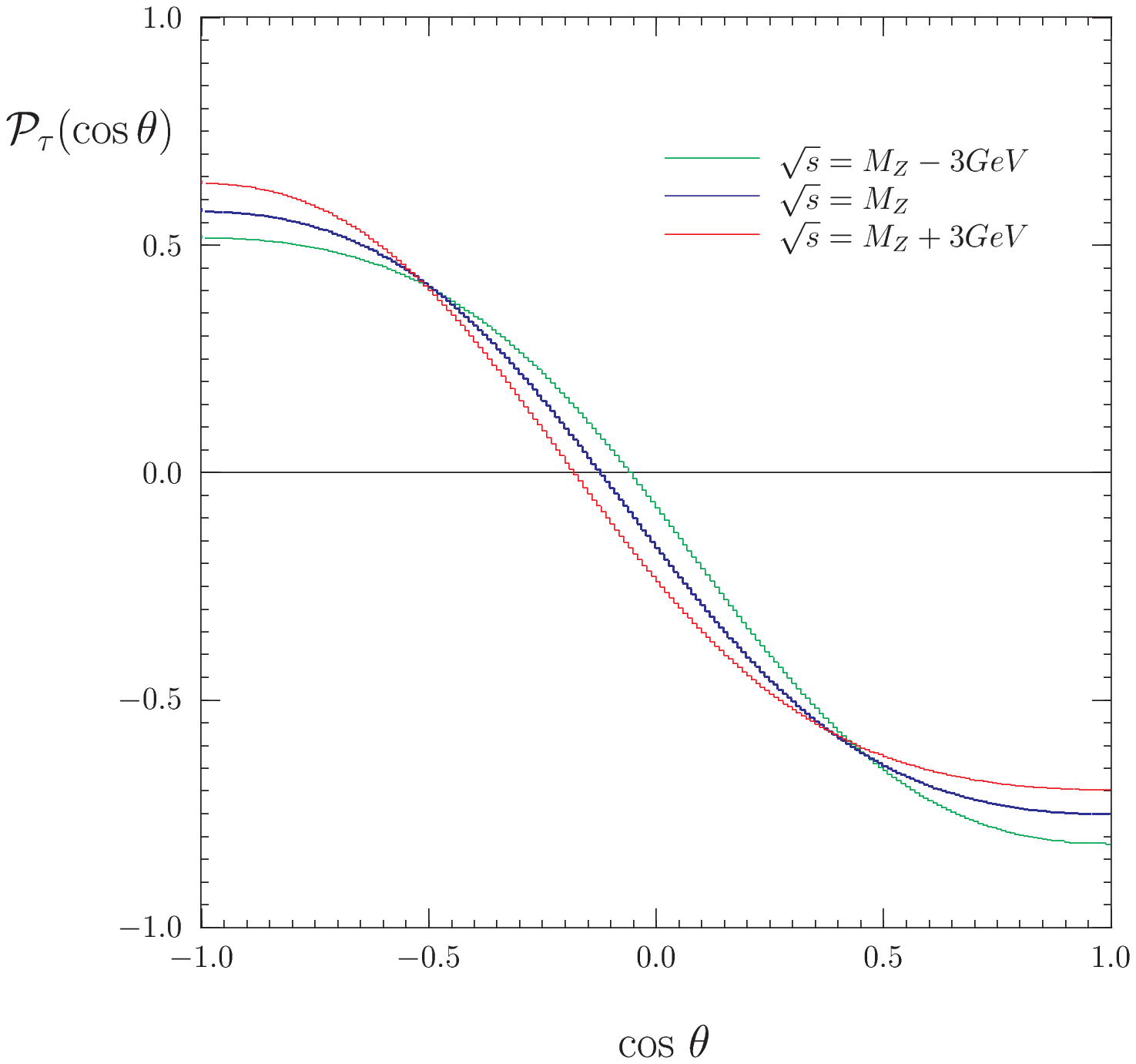,width=110mm,height=130mm}}}
\put( 20,-30){\makebox(0,0)[lb]{\epsfig{file=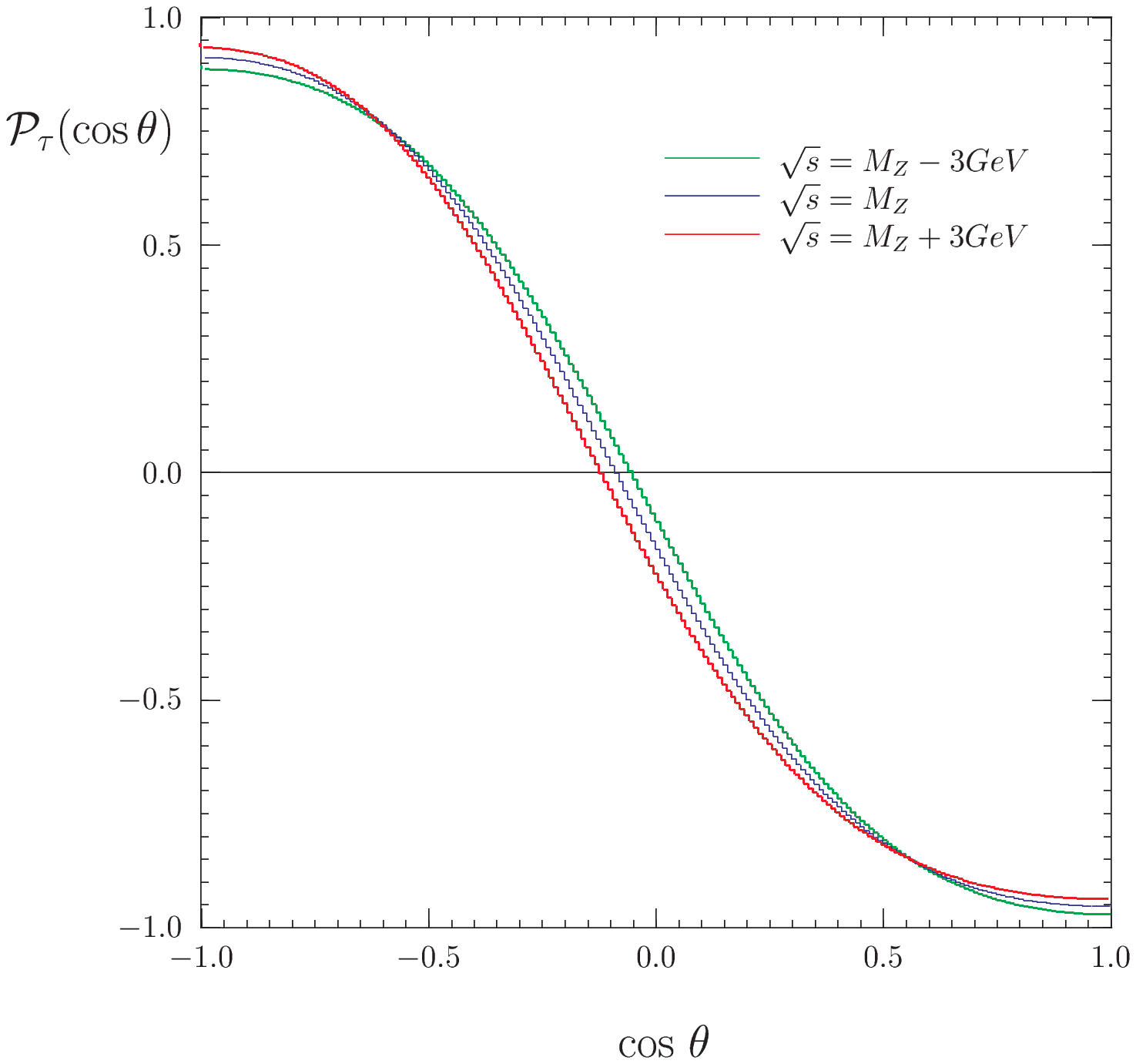,width=110mm,height=130mm}}}
\end{picture}
\end{center}
\caption{Angular dependence of the $\tau$ polarization produced  
through $Z/\gamma$ intermediate state at energies close to the
$Z$ mass from $u$ flavour (left side) and $d$ flavour (right side).
See ref.\cite{Pierzchala:2001gc} for details. }
\label{fig4}
\end{figure}

As we can see in figs. \ref{fig3} and \ref{fig4}, all leading spin effects are nicely 
reproduced by the above set of programs. A more complete discussion
can be found in ref.\cite{Pierzchala:2001gc}.

%-----------------------------------------------------------
%%%%%%%%%%%%%%%%%%%%%%%%%%%%%
%%%%%%%%%%%%%%%%%%%%%%%%%%%%%
%-----------------------------------------------------------

%-----------------------------------------------------------
%%%%%%%%%%%%%%%%%%%%%%%%%%%%%
%%%%%%%%%%%%%%%%%%%%%%%%%%%%%
%-----------------------------------------------------------

The    Monte Carlo {\tt PHOTOS},  
is another example of a program that uses a
{\tt HEPEVT } common block as a data structure. Its role is to generate, whenever suitable,
bremsstrahlung photons in the decays of particles, resonances or sometimes from 
other intermediate charged states.
The program 
 was developped starting from the careful downgrading of the matrix element  
{\tt MUSTRAAL} Monte Carlo \cite{mustraal-cpc:1983}, more precisely its  part 
describing the decay of a $Z$ to a pair of leptons.
Then, the algorithm was
extended to work for the decay of ``any'' particle or resonance.
By construction  it is limited to  leading logarithmic  (ll) approximation with proper
soft-photon angular distributions only.
Thanks to comparisons with codes based on matrix elements, 
{\tt PHOTOS} was checked to performed better than ll
in  the following cases:
 $\tau \to e \nu \bar \nu \gamma$, $\tau \to \pi \nu \gamma$, 
$Z \to \mu^+ \mu^- \gamma (\gamma)$, $g g \to t \bar t \gamma (\gamma)$.

\begin{figure}[!ht]
%\centering
\setlength{\unitlength}{0.1mm}
\begin{picture}(1600,800)
%\put( 0,0){\framebox( 1600,800){ }}
\put( 375,750){\makebox(0,0)[b]{\large }}
\put(1225,750){\makebox(0,0)[b]{\large }}
\put(-300, -400){\makebox(0,0)[lb]{\epsfig{file=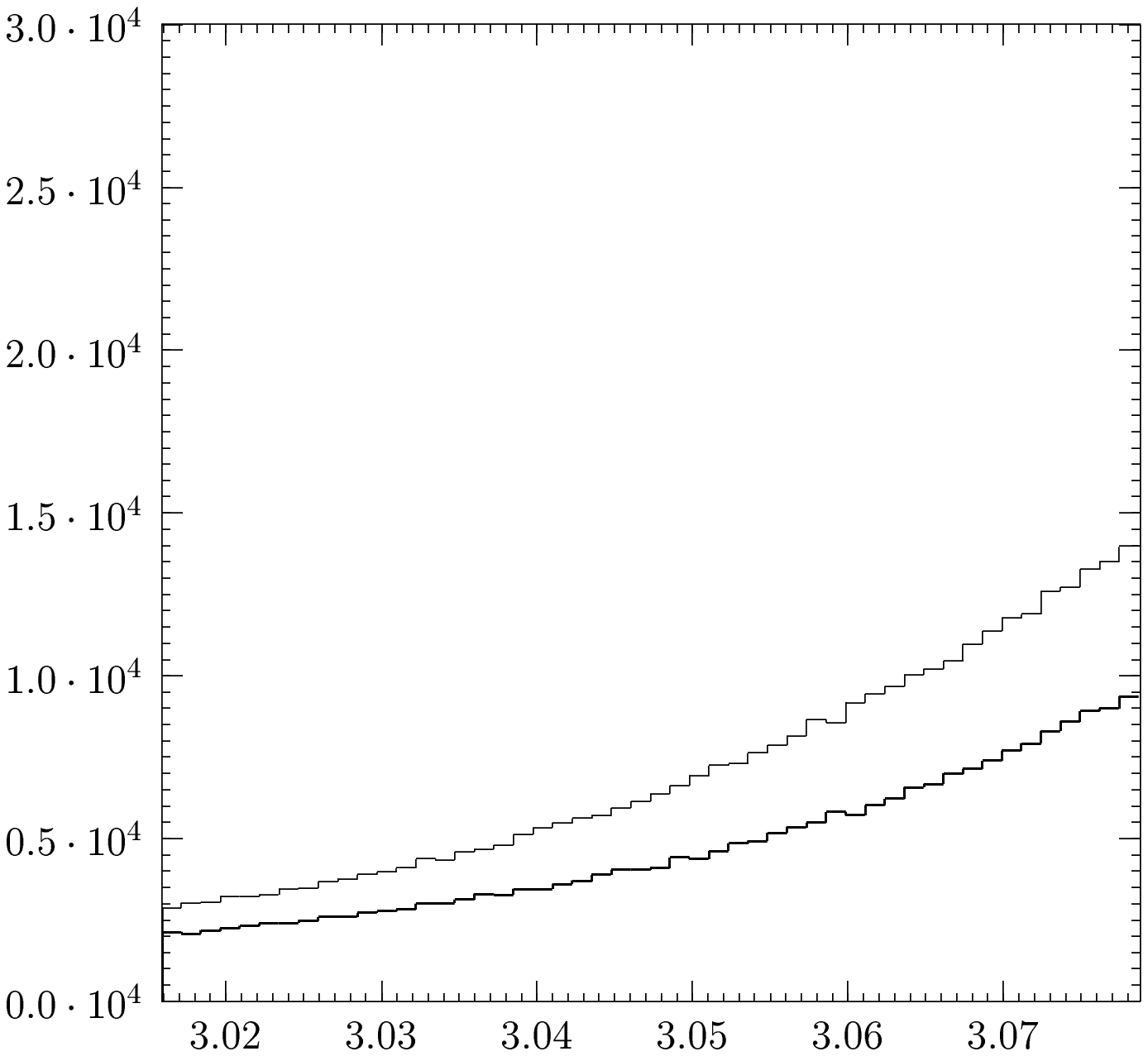,width=120mm,height=140mm}}}
\put(500, -400){\makebox(0,0)[lb]{\epsfig{file=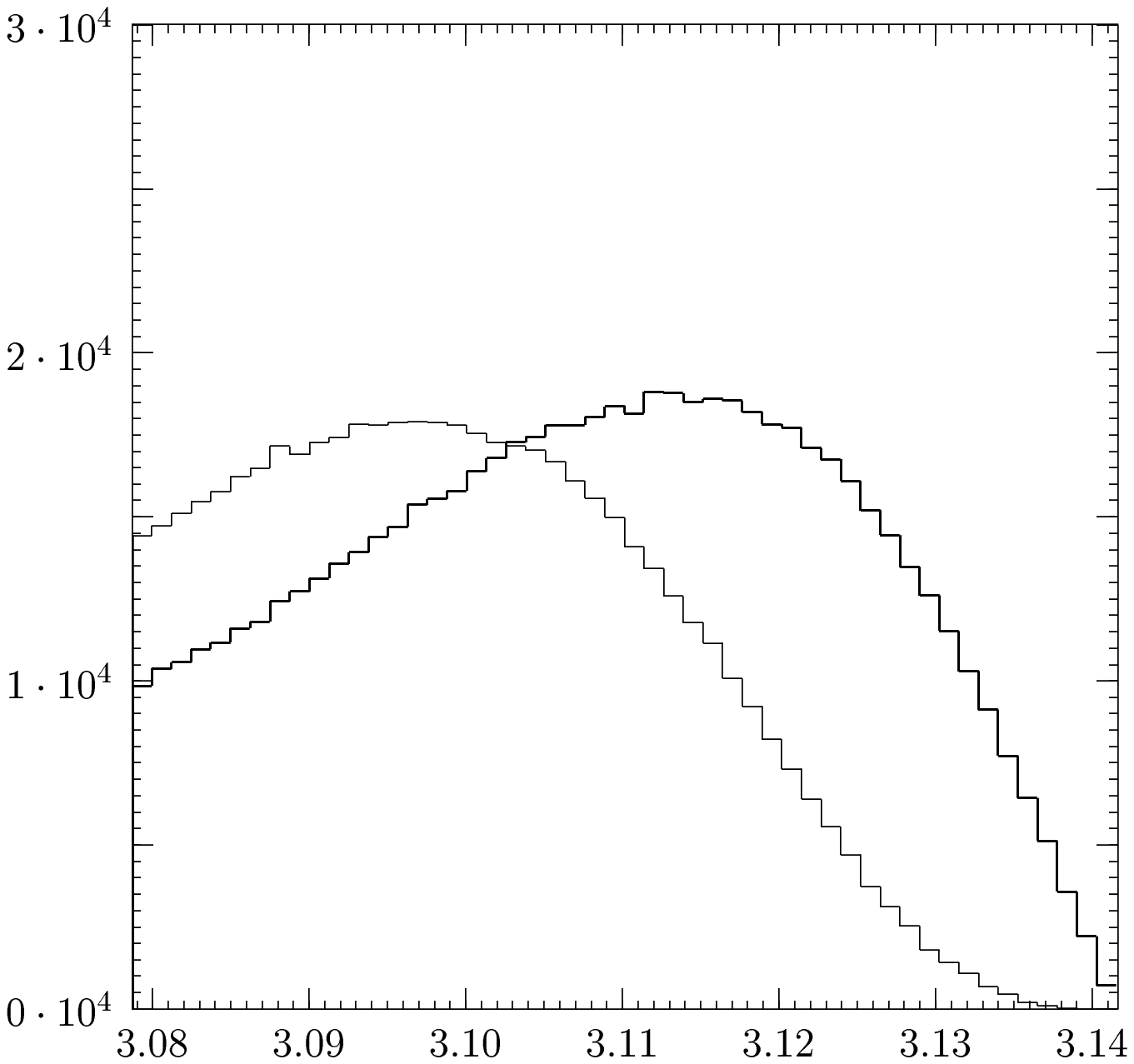,width=120mm,height=140mm}}}
\end{picture}
\caption
{ The  $\pi^+ \pi^-$ acollinearity distribution  (angle $\delta^*$)  
in the  Higgs boson rest frame.
Parts of the distribution close to the end of the spectrum; $\delta^* \sim \pi$ are shown.
The thick line denotes the case 
of the scalar Higgs boson and 
the thin line the pseudoscalar one.}
\label{fig5}
\end{figure}

\begin{figure}[!ht]
%\centering
\setlength{\unitlength}{0.1mm}
\begin{picture}(1600,800)
%\put( 0,0){\framebox( 1600,800){ }}
\put( 375,750){\makebox(0,0)[b]{\large }}
\put(1225,750){\makebox(0,0)[b]{\large }}
\put(-300, -400){\makebox(0,0)[lb]{\epsfig{file=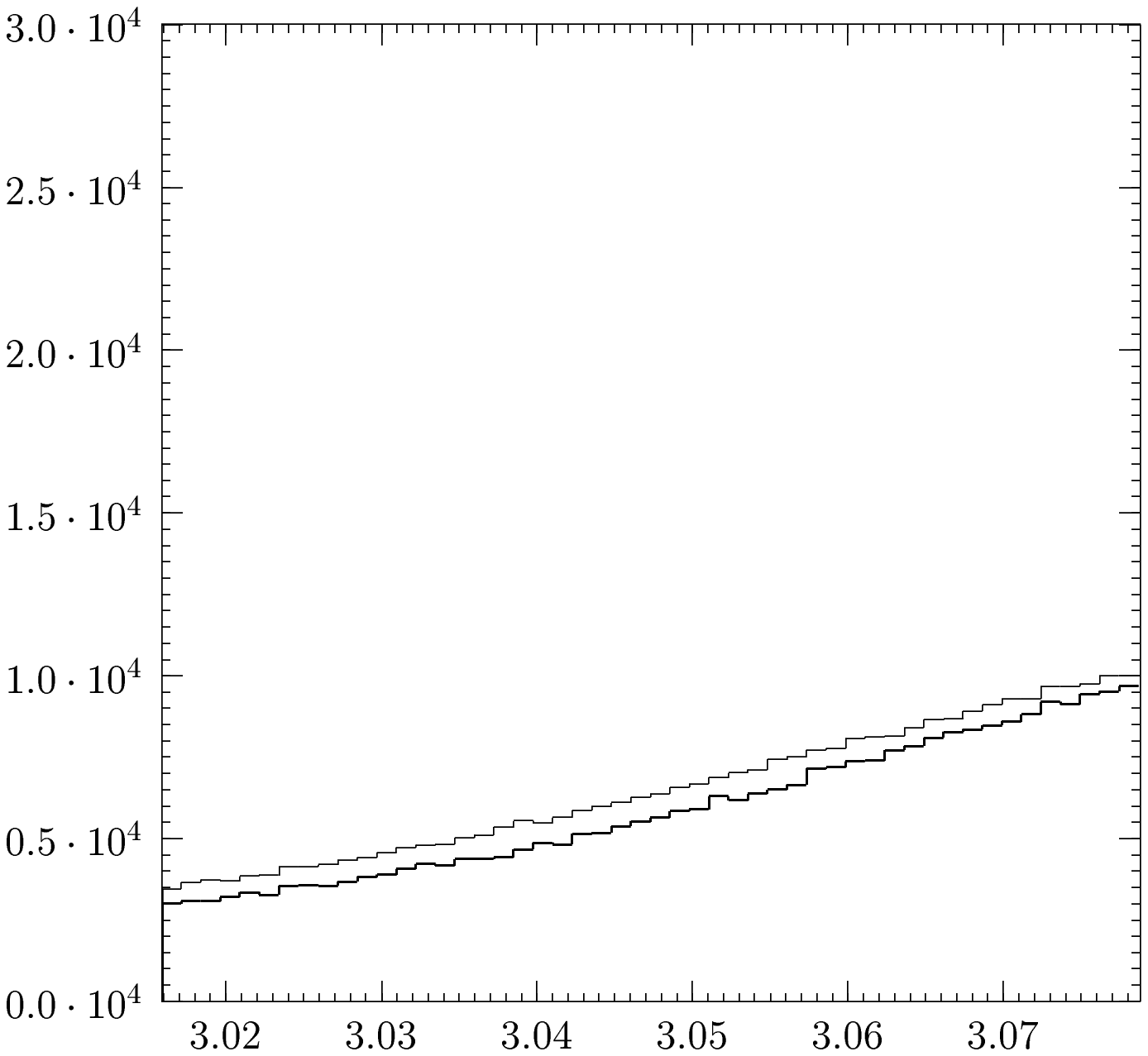,width=120mm,height=140mm}}}
\put( 500, -400){\makebox(0,0)[lb]{\epsfig{file=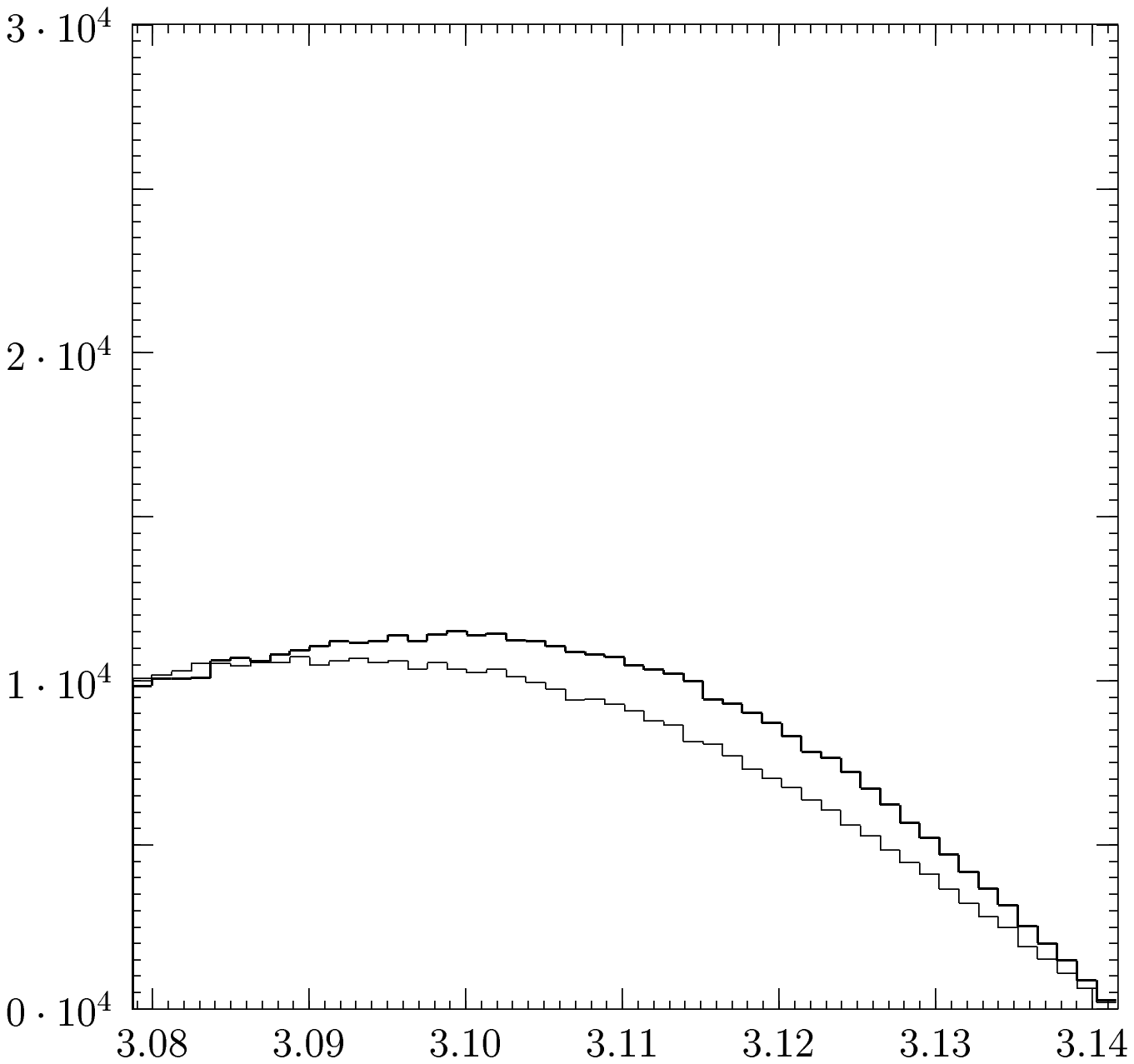,width=120mm,height=140mm}}}
\end{picture}
\caption
{ The  $\pi^+ \pi^-$ acollinearity distribution  (angle $\delta^*$)  
in the scalar Higgs boson reconstructed rest frame. 
Parts of the distribution close to the end of the spectrum; $\delta^* \sim \pi$ are shown.
The thick line denotes the case when all spin effects are included, while  only longitudinal spin correlations are taken for 
the thin line.
 }
\label{fig6}
\end{figure}

In the special case  of a hypothetical (scalar or pseudoscalar)  Higgs boson decay 
$h \to \tau^+\tau^-$, it is possible to define
the full spin density matrix for the pair of $\tau$-leptons, independently of the
Higgs boson production mechanism. The previously discussed algorithm of 
the spin implementation was
extended  (ref.~\cite{Was:2002gv}) to include the full spin effects in that case.
As we can see in fig.~\ref{fig5}, the program  reproduces the effects of the Higgs parity 
 correctly for the $\pi^+$   $\pi^-$ acollinearity calculated  
in the rest frame of the Higgs boson
and decay chain $h \to \tau^+\tau^-;\; \tau^\pm\to \pi^\pm \nu_\tau$ \cite{Kramer:1994jn}.  When some 
detector smearings are  introduced, see fig. \ref{fig6}, the effect  becomes  
less visible.  In this example,
the Higgs-strahlung production mechanism at 350 GeV centre-of-mass energy was 
generated with the help of {\tt PYTHIA} \cite{pythia}; see 
ref.~\cite{Was:2002gv} for details of the study.

\vskip 3mm

\centerline{\large \bf Summary}

\vskip 3mm

We can conclude that, in none of the discussed cases was it necessary to store 
spin degrees of freedom 
in the event records. In fact, storing such information in an approximate way as an attribute 
of particles would not lead to the correct solutions anyway. The 
general principle was shown of how to construct an interface 
that can calculate the relevant multiparticle
density matrix from the kinematical information stored in the event record. 
A backup solution was always necessary. 

Another solution relies on algorithms where decays and productions of some
intermediate states are embodied in the monolithic program individually tailored
to the process. 
This is a good approach in cases such as production and decay of 
$(\tau\tau)\; (WW)\;  (ZZ)$ intermediate states. 
It may be reasonable as well for the case 
of $t\bar t$ production and decay, even though at least 6-body 
final states will have to be generated in single steps. Such a solution
may pose problems if production  and decay
of pairs of {\it new} particles carrying spin will be discovered, 
especially if a multitude of decay 
channels, each described by a distinct model, would have to be combined.

We think that non-factorizability of the spin density matrices into properties
of individual particles is important, and should be borne in mind 
in the discussions of future standards for event records, and
redesigned software for high energy physics.

%*******************************************************************
%*************************************************************************

%*******************************************************************
%*************************************************************************

\vskip 3 mm
\centerline{\large \bf Acknowledgements}
\vskip 3 mm
The inspiring atmosphere at the ECFA-DESY workshop in Cracow and CPP Tokyo 
was essential to the performance of  this work. The
author is specially grateful to M. Ronan  for a discussion.
 
%\bibliographystyle{utphys_spires}
%\bibliographystyle{plain}
%\bibliography{TAUOLA-F}
\providecommand{\href}[2]{#2}
                                     %%%%%%%%%%%%%%%%%%
\end{document}